\RequirePackage{fix-cm}
\documentclass[twocolumn,epjc3]{svjour3}  
\smartqed  % flush right qed marks, e.g. at end of proof
\RequirePackage{graphicx}
\RequirePackage[colorlinks,citecolor=blue,urlcolor=blue,linkcolor=blue]{hyperref}
\RequirePackage{amssymb}
\RequirePackage{amsmath}
\RequirePackage{booktabs}
\RequirePackage{multirow}
\RequirePackage{subcaption}
\RequirePackage{upgreek}
\RequirePackage{xspace}

\newcommand{\decay}[2]{\ensuremath{#1\!\to #2}\xspace} 
\newcommand{\order}[0]{{\ensuremath{\mathcal{O}}}\xspace}

\newcommand{\Pb}[0]{\ensuremath{b}\xspace}           
\newcommand{\Pc}[0]{\ensuremath{\mathrm{c}}\xspace}   
\newcommand{\Pp}[0]{\ensuremath{\mathrm{p}}\xspace}
\newcommand{\PB}[0]{\ensuremath{B}\xspace}                
\def\PD      {\ensuremath{\mathrm{D}}\xspace}    
\newcommand{\PJ}[0]{\ensuremath{J}\xspace}
\newcommand{\PK}[0]{\ensuremath{K}\xspace}

\newcommand{\Pmu}[0]{\ensuremath{\mu}\xspace}
\newcommand{\Ppi}[0]{\ensuremath{\uppi}\xspace}
\newcommand{\Ppsi}[0]{\ensuremath{\uppsi}\xspace}   

\newcommand{\pion}[0]{{\ensuremath{\Ppi}}\xspace}

\newcommand{\pip}[0]{{\ensuremath{\pion^+}}\xspace}
\newcommand{\pim}[0]{{\ensuremath{\pion^-}}\xspace}

\newcommand{\kaon}[0]{{\ensuremath{\PK}}\xspace}
\newcommand{\Kp}[0]{{\ensuremath{\kaon^+}}\xspace}

\newcommand{\proton}[0]{{\ensuremath{\Pp}}\xspace}

\newcommand{\cquark}[0]{{\ensuremath{\Pc}}\xspace}
\newcommand{\bquark}[0]{{\ensuremath{\Pb}}\xspace}

\newcommand{\mup}[0]{{\ensuremath{\Pmu^+}}\xspace}
\newcommand{\mun}[0]{{\ensuremath{\Pmu^-}}\xspace}

\newcommand{\jpsi}[0]{{\ensuremath{{\PJ\mskip -3mu/\mskip -2mu\Ppsi}}}\xspace}

\newcommand{\B}[0]{{\ensuremath{\PB}}\xspace}
\newcommand{\Bu}[0]{{\ensuremath{\B^+}}\xspace}

\def\Bd      {{\ensuremath{\B^0}}\xspace}

\newcommand{\D}[0]{{\ensuremath{\PD}}\xspace}
\newcommand{\Dm}[0]{{\ensuremath{\D^-}}\xspace}

\newcommand{\aunit}[1]{\ensuremath{\text{\,#1}}}

\newcommand{\cm}[0]{\ensuremath{\aunit{cm}}\xspace}

\def\mhz  {\ensuremath{\aunit{MHz}}\xspace}
\def\khz  {\ensuremath{\aunit{kHz}}\xspace}

\newcommand{\ns}[0]{\ensuremath{\aunit{ns}}\xspace}
\newcommand{\ps}[0]{\ensuremath{\aunit{ps}}\xspace}

\newcommand{\gev}{\aunit{Ge\kern -0.1em V}\xspace}
\newcommand{\gevc}{\ensuremath{\aunit{Ge\kern -0.1em V\!/}c}\xspace}

\newcommand{\mev}{\aunit{Me\kern -0.1em V}\xspace}
\newcommand{\mevc}[0]{\ensuremath{\aunit{Me\kern -0.1em V\!/}c}\xspace}
\newcommand{\mevcc}[0]{\ensuremath{\aunit{Me\kern -0.1em V\!/}c^2}\xspace}

\newcommand{\pt}[0]{\ensuremath{p_{\mathrm{T}}}\xspace}

\newcommand{\lhcb}[0]{\mbox{LHCb}\xspace}
\def\lhc    {\mbox{LHC}\xspace}

\newcommand{\hltone}[0]{\texttt{HLT1}\xspace}
\newcommand{\hlttwo}[0]{\texttt{HLT2}\xspace}

\newcommand{\pythia}[0]{\mbox{\texttt{Pythia}}\xspace}
\newcommand{\evtgen}[0]{\mbox{\texttt{EvtGen}}\xspace}
\newcommand{\photos}[0]{\mbox{\texttt{Photos}}\xspace}
\newcommand{\geant}[0]{\mbox{\texttt{Geant4}}\xspace}

\newcommand{\pp}[0]{{{\proton}{\proton}}\xspace}
\newcommand{\BdToDPi}[0]{\decay{\Bd}{\Dm(\Kp\pim\pim)\pip}}
\newcommand{\BuToJpsimmK}[0]{\decay{\Bu}{\jpsi(\mup\mun)\Kp}}

\newcommand{\eg}{\mbox{\itshape e.g.}\xspace}
\newcommand{\ie}{\mbox{\itshape i.e.}\xspace}

\RequirePackage{xcolor}
\RequirePackage{pifont}

\definecolor{tickGreen}{RGB}{74, 195, 65}
\definecolor{queryYellow}{RGB}{255, 200, 01}
\definecolor{crossRed}{RGB}{215, 38, 61}

\def\tick {{\color{tickGreen}{\ding{51}}}\xspace}
\def\query {{\color{queryYellow}{\ensuremath{\boldsymbol{\sim}}}}\xspace}
\def\cross {{\color{crossRed}{\ding{55}}}\xspace}

\journalname{Eur. Phys. J. C}
\begin{document}

\title{Decorrelation of neural networks from particle lifetimes in the LHCb topological $\boldsymbol{b}$ trigger%\thanksref{t1}
}
% \subtitle{Do you have a subtitle?\\ If so, write it here}


\author{
Johannes Albrecht\thanksref{addr1} \and
Alessandro Bertolin\thanksref{addr2} \and
James Connaughton\thanksref{addr3} \and
Jonathan Davies\thanksref{addr4,f1} \and
Blaise Delaney\thanksref{addr6,f2} \and
Agnieszka Dziurda\thanksref{addr7} \and
Conor Fitzpatrick\thanksref{addr5} \and
Maciej Giza\thanksref{addr7} \and
Vava V. Gligorov\thanksref{addr8} \and
James A. Gooding\thanksref{addr1,e1} \and
Nicole Schulte\thanksref{addr1,e2,f3} \and
Nicole Skidmore\thanksref{addr3} \and
Mika Vesterinen\thanksref{addr3} \and
Mike Williams\thanksref{addr10,addr11} \and
Shunan Zhang\thanksref{addr12} \and
Valeriia Zhovkovska\thanksref{addr3}
}

% \thankstext{t1}{Grants or other notes
\thankstext{e1}{Corresponding author: jamie.gooding@cern.ch}
\thankstext{e2}{Corresponding author: nicole.schulte@tu-dortmund.de}
\thankstext{f1}{Formerly ${}^5$}
\thankstext{f2}{Formerly ${}^{10,11}$}
\thankstext{f3}{Formerly ${}^{1}$}

\institute{Fakult{\"a}t Physik, Technische Universit{\"a}t Dortmund, Dortmund, Germany \label{addr1}
\and
    INFN Sezione di Padova, Padova, Italy \label{addr2}
\and
    Department of Physics, University of Warwick, Coventry, United Kingdom \label{addr3}
\and
    Department for Work and Pensions, Manchester, United Kingdom \label{addr4}
\and
    Department of Physics and Astronomy, University of Manchester, Manchester, United Kingdom \label{addr5}
\and
    TimeTrace Labs, London, United Kingdom \label{addr6}
\and
    Henryk Niewodniczanski Institute of Nuclear Physics Polish Academy of Sciences, Krak{\'o}w, Poland \label{addr7}
\and
    Laboratoire de Physique Nucl{\'e}aire et de Hautes {\'E}nergies (LPNHE), Sorbonne Universit{\'e}, CNRS/IN2P3, Paris, France \label{addr8}
\and
    Foodforecast Technologies, Cologne \label{addr9}
\and
    Massachusetts Institute of Technology, Cambridge, MA, United States \label{addr10}
\and
    NSF AI Institute for Artificial Intelligence and Fundamental Interactions, Cambridge, MA, United States \label{addr11}
\and
    Department of Physics, University of Oxford, Oxford, United Kingdom \label{addr12}
}

\date{Submitted to Eur.~Phys.~J.~C: 24${}^{\rm th}$ July 2026}

\maketitle

\begin{abstract}
    The \lhcb topological beauty trigger is the primary set of algorithms for selecting collision events containing \bquark-hadrons in the fully software-based \lhcb trigger.
    The algorithms apply monotonic Lipschitz neural networks (NNs) to select vertices of charged particles consistent with the distinct topology of a \bquark decay, \ie, those with large lifetimes and transverse momentum.
    Many analyses of the events recorded require that the selection must be unbiased with respect to the \bquark-hadron lifetime at large lifetimes.
    Accurate reconstruction is challenging in busier detector environments, in which several visible proton-proton collisions occur simultaneously per bunch crossing, such that mis-association of decay products can result in vertices with artificially large measured lifetimes.
    This paper presents two approaches to mitigate correlations between NN scores and candidate lifetimes at large lifetime, and evaluates the performance of the resulting models.
% \keywords{First keyword \and Second keyword \and More}
\end{abstract}

\section{Introduction}
\label{sec:introduction}

The topological beauty (\bquark) trigger of the Large Hadron Collider beauty (\lhcb) experiment \cite{LHCb-DP-2008-001} is a machine learning (ML)-based selection algorithm which identifies reconstructed candidates, \ie, events containing physics of interest, consistent with the decay topology of hadrons containing \bquark-quarks \cite{LHCb-PUB-2011-002,LHCb-PUB-2011-016}.
This trigger is part of the second stage of the software-based High Level Trigger (HLT) and provides the largest sample of events containing \bquark-hadrons of the HLT algorithms \cite{LHCb-DP-2019-001}.
In particular, the topological \bquark trigger is commonly used as an input to time-dependent measurements.
The trigger response must thus be unbiased with respect to particle lifetime at large lifetimes.
In Run 3 of the Large Hadron Collider (LHC), this has been made more challenging by an increase in the number of visible proton-proton (\pp) collisions per bunch crossing, $\mu$, from $\mu=1.1$ in Runs 1 and 2 to $\mu=5.3$ in Run 3.

The implementation of the topological \bquark trigger in Run 3 of the \lhc using monotonic Lipschitz neural networks (MLNNs) \cite{lipschitz-paper,topo-acat-2022,topo-chep-2023} is described in Sec.~\ref{sec:topo}.
In Sec.~\ref{sec:background}, the challenge presented by multiple \pp collisions per bunch crossing is described, and a specific lifetime-correlated background contribution is identified.
Methods of decorrelating the MLNNs with respect to candidate lifetime for large lifetimes are presented in Sec.~\ref{sec:decorrelation} and demonstrated in the context of the topological \bquark trigger in Sec.~\ref{sec:performance}.

\section{The topological \bquark trigger in Run 3}
\label{sec:topo}

The \lhcb experiment \cite{LHCb-DP-2008-001} at the \lhc is a single-arm forward spectrometer,  optimised for the study of heavy-flavour hadrons and primarily instrumented in the pseudorapidity range $2 < \eta < 5$.
Between Runs 2 and 3 of the \lhc, the experiment was upgraded and the \lhcb trigger was redesigned to remove the hardware-based Level 0 trigger~\cite{LHCb-DP-2022-002}.
This enabled real-time processing of \pp collisions at 30\mhz by the HLT~\cite{LHCb-DP-2019-001,LHCb-DP-2019-002}: the first stage, \hltone, which operates on Graphical Processing Units, partially reconstructs events and reduces the event rate to $\sim1\mhz$ \cite{Aaij:2019zbu}; \hlttwo, which operates on Central Processing Units, fully reconstructs events, reducing the event rate further to $\sim100\khz$.

Reconstructed candidates can be selected either exclusively, targeting a fully specified process, or inclusively, targeting a generic signature.
Exclusively selected events typically save only the candidates reconstructed by the selection algorithm which has fired, per the Turbo event model \cite{LHCb-PROC-2015-011}.
For inclusively selected events, the entire event is usually saved, allowing for further offline processing and reprocessing \cite{Mathad:2023zky,Abdelmotteleb_2025}.

As \bquark-hadrons decay primarily by the weak, CKM-suppressed decay of the \bquark-quark, these hadrons have lifetimes of $\order({1 \ps})$.
At the energies studied by \lhcb, this corresponds to a flight distance of $\order({1\cm})$ traversed by \bquark-hadrons before they decay.
Therefore, the \bquark-hadron decay topology contains a secondary vertex (SV) significantly detached from the primary vertex (PV) at which the \bquark-hadron is produced.
The decay products of the \bquark-hadron originate from the SV, in some cases undergoing further decay processes.
These decay products are used to reconstruct the SV, which can then be characterised by features such as the \bquark-hadron flight distance (distance from PV to SV) and impact parameter (distance of closest approach to the beam axis), which can be used to inclusively select decays consistent with a \bquark-hadron topology.
Since \cquark-hadrons decay with similar topologies, a modified invariant mass accounting for missing transverse momentum can also be used to distinguish between \bquark-/\cquark-hadron decays~\cite{SLD:1997ihw}.
A topological \bquark trigger has been employed in \hlttwo since Run 1, applying ML algorithms to select events consistent with a \bquark-hadron decay based on these features \cite{LHCb-PUB-2011-002,LHCb-DP-2019-001}.
A full list of these features is given in Appendix \ref{app:features}.

The topological \bquark trigger was originally implemented using bonsai boosted decision trees \cite{BBDT,LHCb-PROC-2015-018}; however, ahead of Run 3, these algorithms were replaced with MLNNs to select 2- and 3-body candidates \cite{lipschitz-paper,topo-acat-2022,topo-chep-2023}.
The MLNN architecture was chosen as it provides a monotonic NN response in features of choice.
Specifically, the NNs are required to be monotonic in the minimum transverse momentum, \pt, and impact parameter quality, $\chi^2_{\rm IP}$, of final state particles.
This monotonicity is at the level of the partial derivatives of the NN response in each of the features. 
For example, with all other features held fixed, the NN response must increase with increasing \pt.
Whilst these NNs do not directly access the lifetime of a given \bquark-hadron, as the 2- and 3-body candidates are typically only part of the full \bquark-hadron decay, features such as the $\chi^2_{\rm IP}$ are strongly correlated to the particle lifetime and act as proxies for it.
As a \bquark-hadron with a longer lifetime is more distinct from background contributions, it should be more readily selected and thus the MLNNs should thus return scores which approximately monotonically increase with increasing particle lifetime.
However, it is not possible to explicitly guarantee this formally given that the NNs do not have access to the lifetime itself---and due to the multivariate nature of the problem.

Four selection algorithms are defined, in which 2- and 3-body candidates are reconstructed:
\begin{itemize}
    \item \texttt{Hlt2Topo2Body}: two long tracks, \ie, those with hits in the Vertex Locator \cite{LHCb-TDR-013} and Scintillating Fibre Tracker \cite{LHCb-TDR-015}, are combined to form an SV, detached from their associated PV~\cite{Dziurda:2024xpw}.
    \item \texttt{Hlt2Topo3Body}: a third track is added to the SV, which must be detached from their associated PV.
    \item \texttt{Hlt2TopoMu2Body, Hlt2TopoMu3Body}: identical to their respective n-body counterparts except that one track must leave hits in the \lhcb muon systems as well.
\end{itemize}
The tracks and vertices of each line are subject to an initial set of loose cut-based selection requirements to suppress random combinations of tracks.

Two MLNNs are trained, for 2- and 3-body candidates, respectively, such that looser thresholds can be applied to candidates from the muonic algorithms, as the muon requirement provides a greater suppression of background.
This training is performed in \texttt{PyTorch} \cite{Paszke:2019xhz} by minimising a binary cross-entropy (BCE) loss with the \texttt{Adam} minimiser \cite{Kingma:2014vow}:
\begin{equation}
    \mathcal{L}_{\rm BCE} = -\frac{1}{N} \sum\limits_i y_i \log(\hat{y}_i) + (1 - y_i) \log(1 - \hat{y}_i),
\end{equation}
where $y_i \in \{0, 1\}$ and $\hat{y}_i \in [0, 1]$ are the category label and NN score for event $i$, over the $N$ training events.
The signal category ($y_i = 1$) contains events from a cocktail of simulated decays of interest; the background category ($y_i = 0$) contains minimum bias simulation with any contributions from \bquark hadrons removed.

To produce the simulated samples used in the training, $\proton\proton$ collisions are generated using  \pythia~\cite{Sjostrand:2007gs,Sjostrand:2006za} with a specific \lhcb configuration~\cite{LHCb-PROC-2010-056}.
Decays of unstable particles are described by \evtgen~\cite{Lange:2001uf}, in which final-state radiation is generated using \photos~\cite{davidson2015photos}.
The interaction of the generated particles with the detector, and its response, are implemented using the \geant toolkit~\cite{Allison:2006ve,Agostinelli:2002hh} as described in \cite{LHCb-PROC-2011-006}.

\section{Lifetime-correlated background}
\label{sec:background}

\begin{figure}[t]
    \centering
    \begin{subfigure}[t]{\linewidth}
        \centering
        \includegraphics[width=0.775\linewidth]{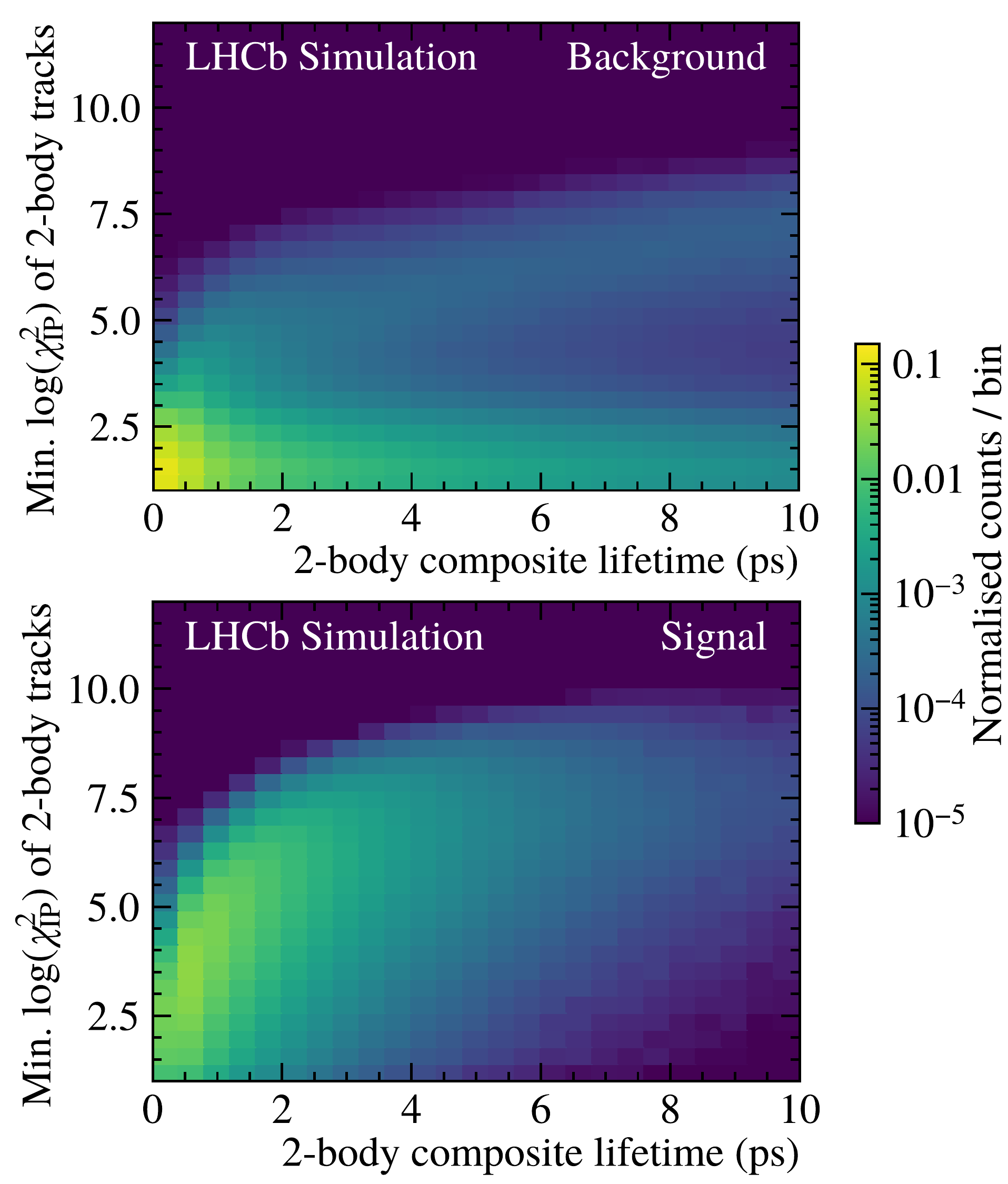}
        \caption{2-body candidates in the (above) background and (below) signal samples.}
        \label{fig:background/two-body}
    \end{subfigure}
    
    \begin{subfigure}[t]{\linewidth}
        \centering
        \includegraphics[width=0.775\linewidth]{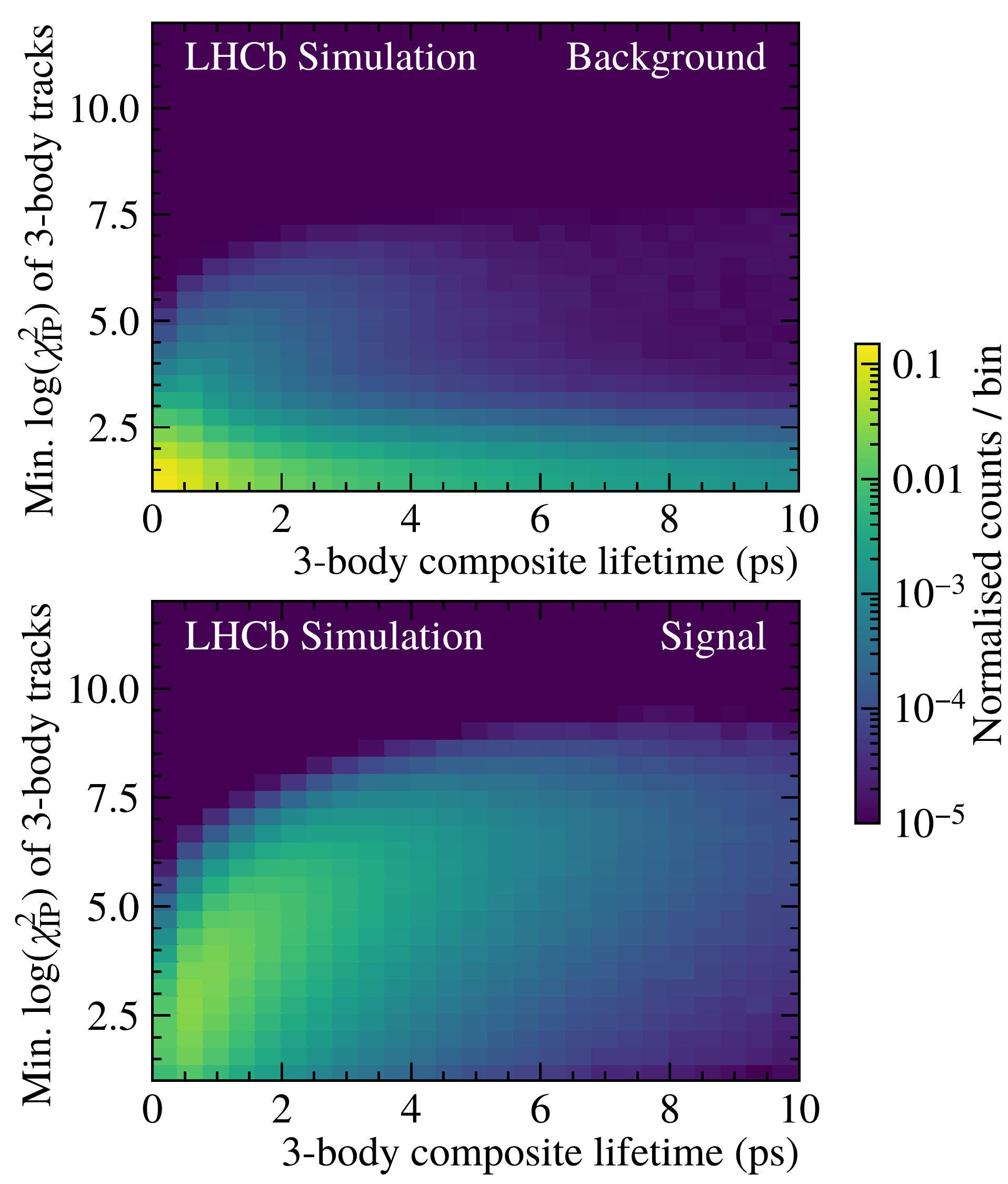}
        \caption{3-body candidates in the (above) background and (below) signal samples.}
        \label{fig:background/three-body_threebody}
    \end{subfigure}
    
    \caption{The composite lifetime and minimum $\log({\chi_{\rm IP}^2})$ of constituent tracks are plotted against one another for 2- and 3-body candidates in minimum bias background and ${\BuToJpsimmK}$ signal training samples.
    Each are normalised according to the number of candidates in the sample.
    }
    \label{fig:background}
\end{figure}

In Runs 1 and 2, \lhcb operated at $\mu=1.1$, meaning that few bunch crossings contained multiple visible \pp collisions.
Therefore, background contributions arose predominantly from particles produced within a single PV.
In raising $\mu$ to $5.3$ in Run 3, backgrounds consisting of particles produced in different PVs became more frequent.
For example, a 3-body candidate may be formed in which the 2-body part consists of tracks from the PV of interest, but the third track is produced in another PV and wrongly associated to the PV of interest.
The resulting SV of this candidate would therefore be highly biased, with a flight distance and subsequent lifetime, $\tau$, which can be comparable in size to vertices produced by the decays of long-lived signals of interest.
To illustrate this, the correlation of composite lifetime and the minimum $\log({\chi^2_{\rm IP}})$ of constituent tracks ($\min(\log\chi^2_{\rm IP\!,\, fs})$) is presented for 2- and 3-body candidates in Fig.~\ref{fig:background} for the minimum bias background and ${\BuToJpsimmK}$ signal training samples.

\begin{figure}[t]
    \centering
    \includegraphics[width=0.9\linewidth]{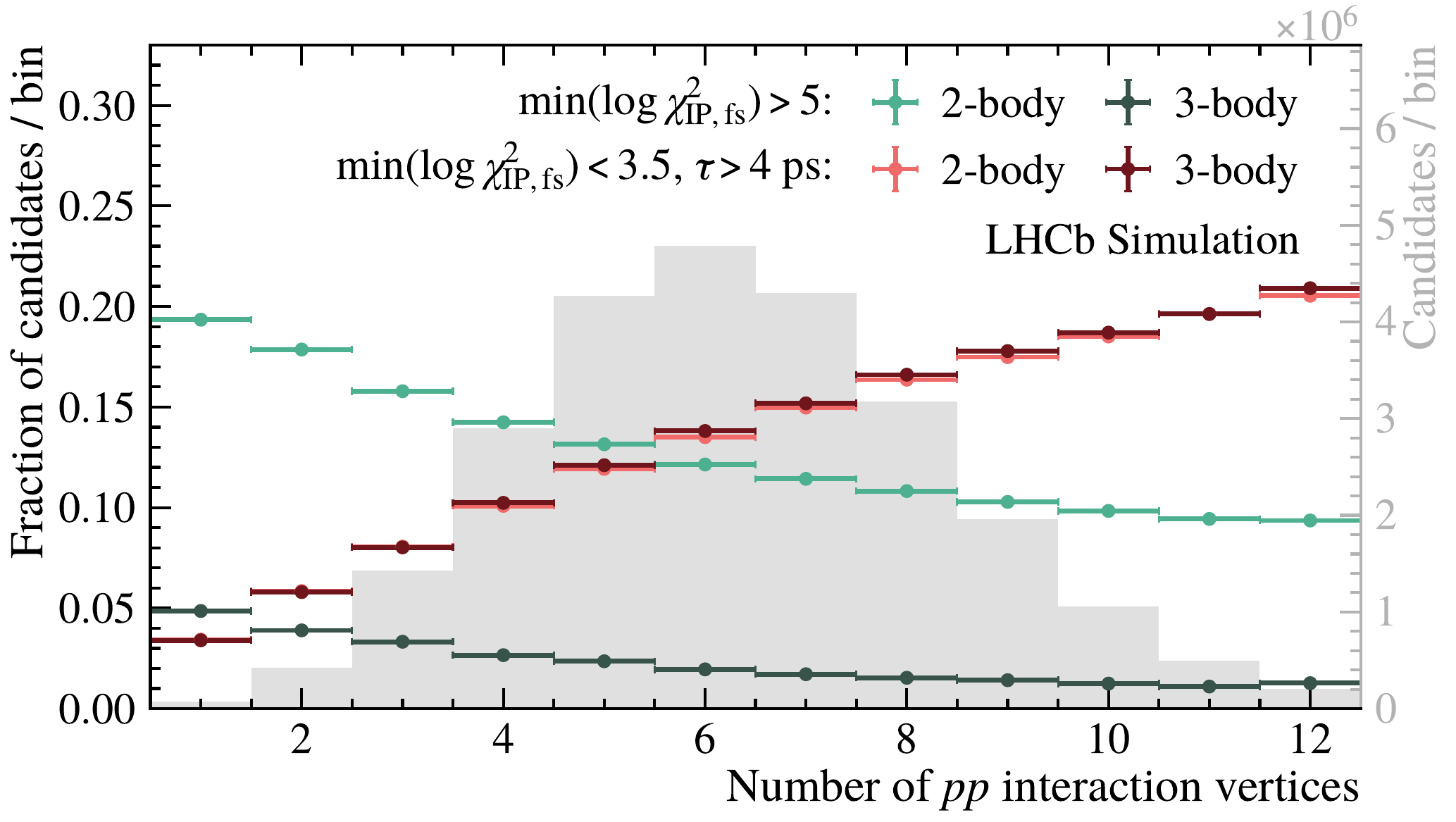}
    \caption{Fraction of 2- and 3-body candidates in the background sample lying within regions dominated by combinatorial and lifetime-persistent backgrounds, evaluated in bins of the number of PVs in the event.
    The total number of candidates is shown as an underlying grey distribution.}
    \label{fig:background-nPVs}
\end{figure}

In both the background and signal samples, a component is seen extending in an arc from the bottom left to top right corners of each plot.
In signal this is the expected distribution of candidates; in background, this is the equivalent contribution from combinatorial background.
However, an additional component is observed for low $\min(\log\chi^2_{\rm IP\!,\, fs})$, \ie, for composites containing at least one PV-associated track which is likely to have originated from a \bquark decay, which extends from $\tau = 0\ps$ up to $\tau = 10 \ps$.
This is precisely the background discussed above, wherein candidates are reconstructed from tracks of different PVs, as is demonstrated by the dependence of this background and the combinatorial background on the number of PVs in the event, shown in Fig.~\ref{fig:background-nPVs}.
For large lifetimes, this becomes the dominant contribution, as both the signal and combinatorial background fall off exponentially.

As a consequence of the background from PV misassociation, the large lifetime region is associated with a drop in signal purity, which can lead a NN to penalise candidates with large lifetimes.
For decay-time-dependent studies, this results in a sparse population of large-lifetime candidates, reducing sensitivity to the $\cosh$ term of the \bquark-hadron decay-time distribution~\cite{dunietz-cpv}.
To prevent this, the loss with which the NNs are trained must be modified to decorrelate the NN response with respect to the candidate lifetime at large decay times.

\section{Decorrelating neural networks}
\label{sec:decorrelation}

The fundamental approach to decorrelating an NN with respect to a given variable, denoted $\tau$ for convenience, is relatively straightforward: a penalty term must be included in the loss function, which is dependent on the correlation of the NN score, $\hat{y}$, with $\tau$.
Many penalty terms have been devised and two such terms are considered here: distance correlation (DisCo) \cite{disco-paper} and moment decomposition (MoDe) \cite{mode-paper}, as these have shown promise in high-energy physics contexts, namely in preventing mass-sculpting in multivariate classifiers.

The DisCo approach defines a penalty term from the distance covariance \cite{distance-correlation} of $\hat{y}$ with $\tau$, normalised by the distance covariances of $\hat{y}$ and $\hat{\tau}$ with themselves:
\begin{equation}
    \mathcal{L} = \mathcal{L}_\mathrm{BCE} + \lambda \cdot \underbrace{
        \frac{
            \mathrm{dCov}(\hat{y}, \tau)
        }{
            \sqrt{
                \mathrm{dCov}(\hat{y}, \hat{y}) \cdot \mathrm{dCov}(\tau, \tau)
            }
        }
    }_{\mathcal{L}_\mathrm{DisCo}\,=\,\mathrm{dCorr}(\hat{y}, \tau)}
\end{equation}
where $\lambda$ is a hyperparameter which parametrises the penalty term strength, such that $\lambda = 0$ returns a BCE loss.
This is applied as linear in $\mathrm{dCorr}(\hat{y}, \tau)$, which maintains a balanced penalty throughout training.

The MoDe approach decomposes the cumulative density function of scores defined in bins of $\tau$ (with central value $\widetilde{\tau}$), $F(\hat{y})_\tau$ into Legendre polynomials of order $\ell$, $\widetilde{F}(\hat{y})^{\ell}_\tau$.
In this basis, the absolute square difference between $F(\hat{y})_\tau$ and $\widetilde{F}(\hat{y})^{\ell}_\tau$ should be minimal for an NN which is uncorrelated to $\tau$ for $\ell = 0$.
Incorporating this penalty term, the loss function becomes
\begin{equation}
    \mathcal{L} = \mathcal{L}_\mathrm{BCE} + \lambda \cdot \underbrace{
        \sum\limits_\tau{\int{
            \left|
                F_\tau(\hat{y}) - \widetilde{F}^\ell_\tau(\hat{y})
            \right|^2~\mathrm{d}\hat{y}
        }} 
    }_{\mathcal{L}_\mathrm{MoDe}^\ell}
\end{equation}
wherein the decomposition $\widetilde{F}(\hat{y})^{\ell}_\tau$ is evaluated as
\begin{equation}
    \widetilde{F}^{\ell}_\tau(\hat{y}) = \sum\limits_{l=0}^\ell{
        c_l(\hat{y}) P_l(\widetilde{\tau})
    }.
\end{equation}
For this paper, $\ell=2$ is used, as a quadratic MoDe term allows for a controlled quadratic dependence between $\tau$ and $\hat{y}$, with the coefficients further restricted to enforce monotonicity in $\tau$.
In Ref.~\cite{mode-paper}, this was seen to provide a superior performance than implementations with $\ell=0,1$. 
This is expected as it allows for (does not penalize) the sharp turn-on curve required at small decay times to reject prompt backgrounds, while still enforcing monotonic behaviour at large decay times.

% \newpage
\section{Performance of lifetime-decorrelated MLNNs}
\label{sec:performance}

The two decorrelation approaches described in Sec.~\ref{sec:decorrelation} were implemented in the training of the MLNNs for the topological \bquark trigger.
Models for selecting 2- and 3-body candidates were trained for a range of decorrelation strengths for each approach, as detailed in Appendix~\ref{app:optimisation}.
For models trained with a DisCo loss term, the strongest decorrelation without a significant reduction in classification power is achieved for strengths of $\lambda = 0.8$ for the 2- and 3-body models.
A strength of $\lambda = 0.2$ was found to be equivalently optimal for both 2- and 3-body models trained with a MoDe term.

To evaluate the impact of each approach on time-dependent physics analyses, simulated samples of reconstructible, \hltone-filtered ${\BuToJpsimmK}$ and ${\BdToDPi}$ decays were processed by \hlttwo, applying each of the models inside \texttt{Hlt2Topo2Body} and \texttt{Hlt2Topo3Body}.
These processes are well-understood and are topologically representative of many decays studied by LHCb, wherein the \bquark-hadron decays via an intermediary which, in the case of ${\BdToDPi}$, flies a measurable distance in the detector before decaying into many products.

\begin{figure}[htb]
    \centering
    \includegraphics[width=0.9\linewidth]{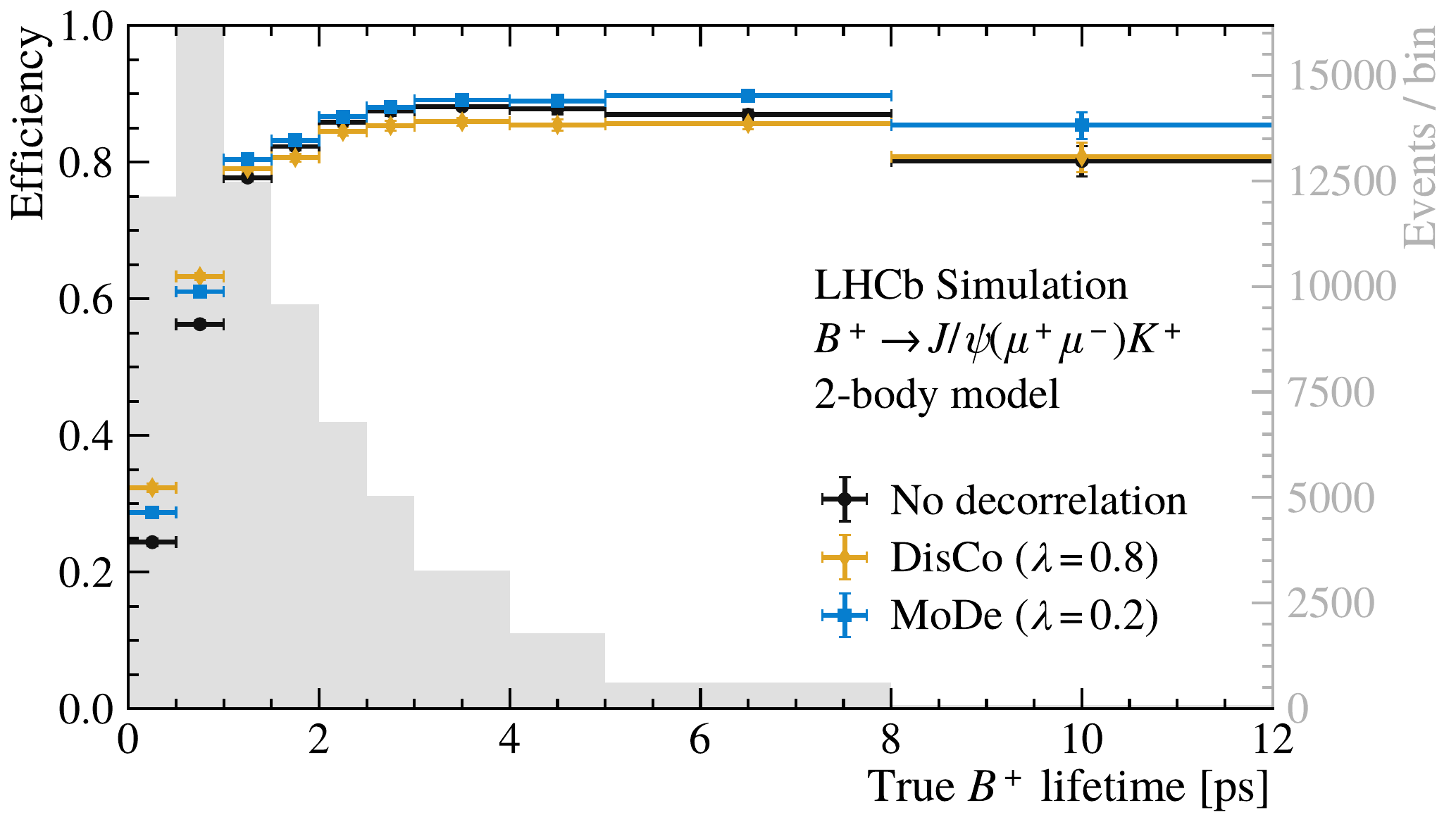}
    \includegraphics[width=0.9\linewidth]{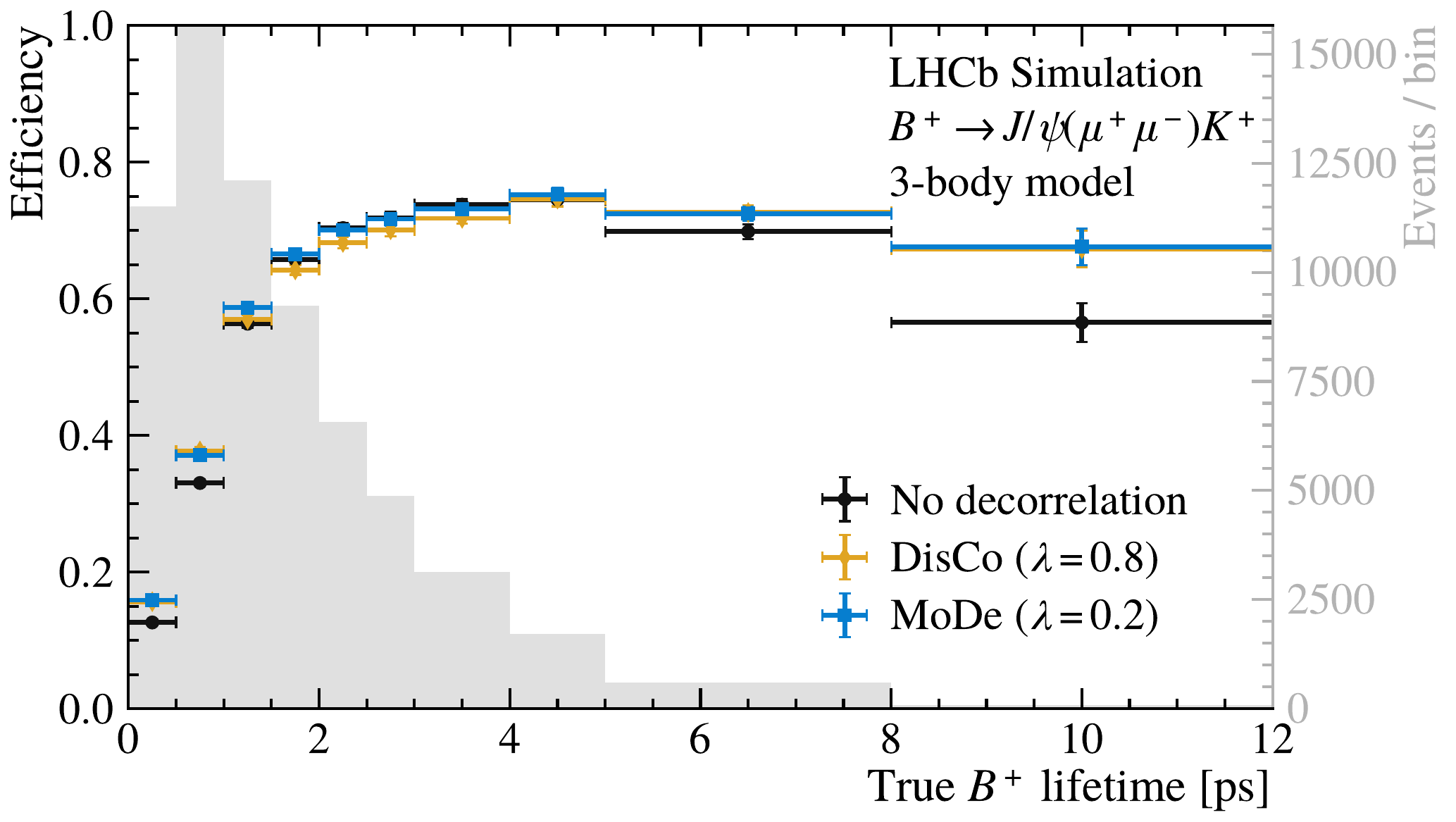}
    \caption{Efficiencies of  (top) 2- and (bottom) 3-body models to select simulated ${\BuToJpsimmK}$ events, which are required to be fully reconstructible and pass both \hltone and the topological \bquark trigger preselection.
    Models are trained with no decorrelation term, a DisCo term of strength $\lambda = 0.8$, and a MoDe term of strength $\lambda = 0.2$.}
    \label{fig:efficiencies-butojpsik}
\end{figure}

\begin{figure}[t]
    \centering
    \includegraphics[width=0.9\linewidth]{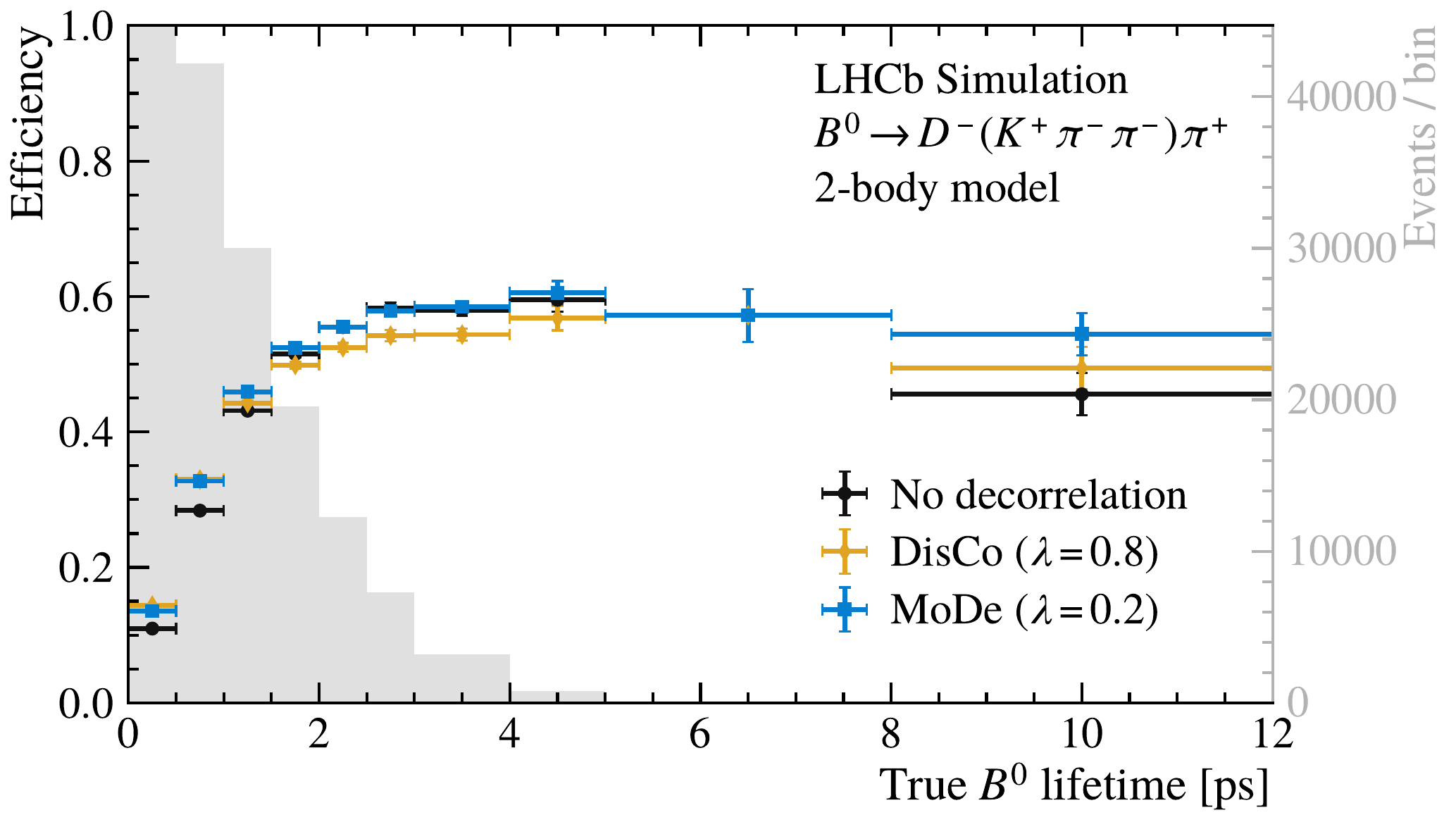}
    \includegraphics[width=0.9\linewidth]{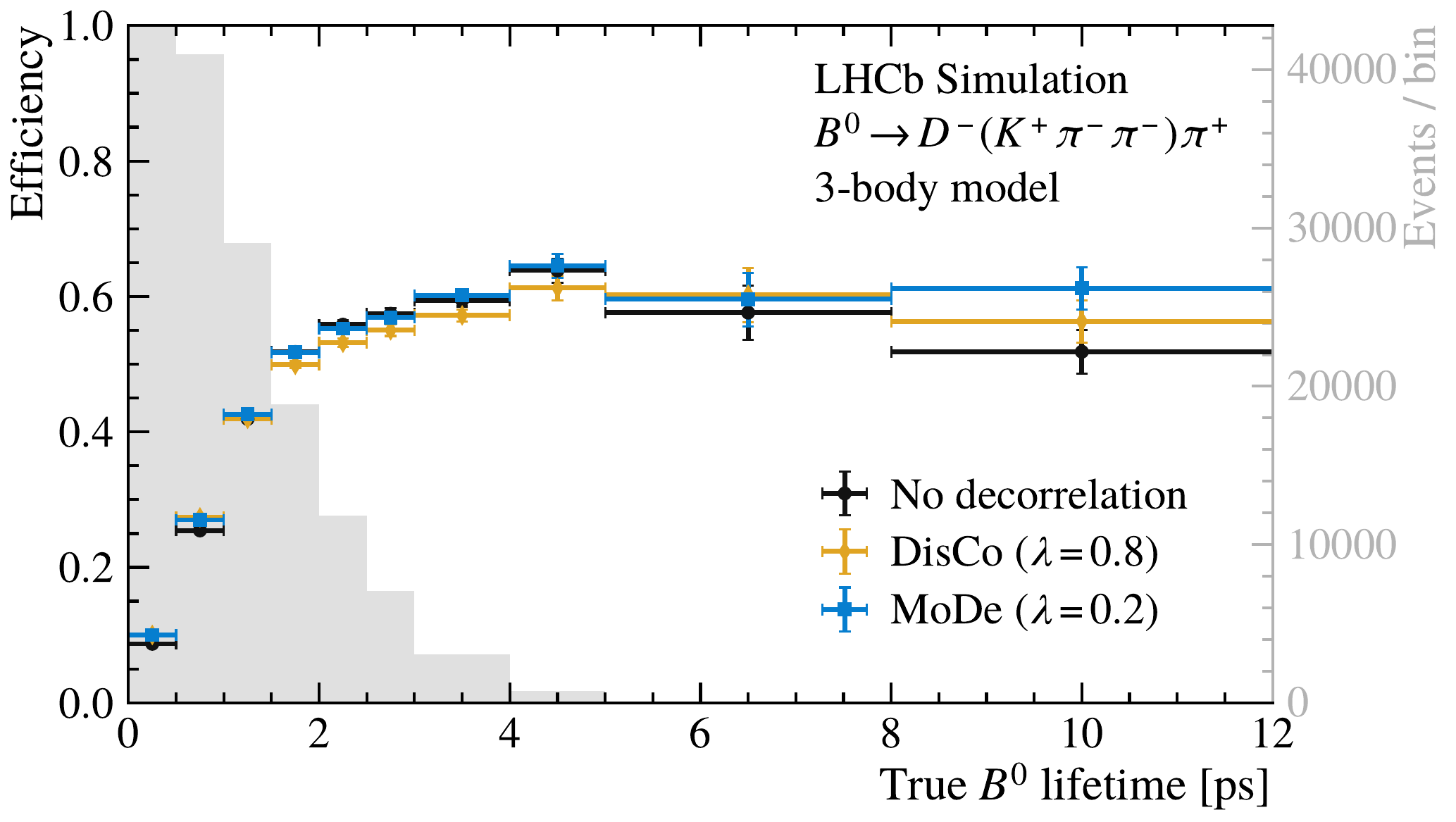}
    \caption{Efficiencies of  (top) 2- and (bottom) 3-body models to select simulated ${\BdToDPi}$ events, which are required to be fully reconstructible and pass both \hltone and the topological \bquark trigger preselection.
    Models are trained with no decorrelation term, a DisCo term of strength $\lambda = 0.8$, and a MoDe term of strength $\lambda = 0.2$.}
    \label{fig:efficiencies-bdtodpi}
\end{figure}

Thresholds are imposed on the output of each model; for the sake of comparison, the thresholds were defined such that \texttt{Hlt2Topo2Body} and \texttt{Hlt2Topo3Body} returned an output bandwidth in line with that typically assigned by \lhcb.
Each \hlttwo algorithm was configured such that the same output bandwidth is returned with each MLNN model applied.
The efficiencies of the resulting selection algorithms, defined as the fraction of pre-selected events which also pass each threshold, were evaluated in the ${\BuToJpsimmK}$ and ${\BdToDPi}$ samples, in bins of true \bquark-hadron lifetime, and are presented in Figs.~\ref{fig:efficiencies-butojpsik}~\&~\ref{fig:efficiencies-bdtodpi}, respectively.

The decorrelated models generally achieve a flatter efficiency at large lifetimes (${\tau \gtrsim 5\ps}$).
For the DisCo approach, this manifests as a reduced efficiency for intermediate lifetimes, which is then maintained into large lifetimes.
For the MoDe approach, this is achieved as an improvement in the large lifetime efficiencies whilst maintaining performance at intermediate lifetimes. 
This is likely due to the version of MoDe used not penalizing the turn-on curve at small decay times. 

To further demonstrate this, linear least-squares regression fits were performed to the efficiencies estimated in ${\BdToDPi}$, over the bins with ${\tau > 4\ps}$, for each decorrelation technique.
The resulting parameters of these fits are listed in Table.~\ref{tab:fits}.
Both decorrelation approaches result in reduced negative slopes, with MoDe achieving a stronger reduction, particularly in the case of 3-body candidates, where the slope is almost consistent with 0.
We thus deem the MoDe approach to be better suited to achieving lifetime-decorrelation in the topological \bquark trigger.

\begin{table*}[htb]
    \centering
    \caption{Parameters resulting from linear least-squares regression fits to the efficiencies shown in Fig.~\ref{fig:efficiencies-bdtodpi}.}
    \begin{tabular}{cccc}
        \toprule
        \multicolumn{2}{c}{Model} & Slope $[\ns^{-1}]$ & Intercept \\
        \midrule
        \multirow{3}{*}{2-body}   & No decorrelation        & $-24.9 \pm 4.2$ & $0.710 \pm 0.026$ \\
                                  & DisCo ($\lambda = 0.8$) & $-12.9 \pm 4.7$ & $0.629 \pm 0.030$ \\
                                  & MoDe ($\lambda = 0.2$)  & $-11.23 \pm 1.6$  & $0.655 \pm 0.010$ \\
        \midrule
        \multirow{3}{*}{3-body}   & No decorrelation        & $-22.2 \pm 2.8$ & $0.736 \pm 0.017$  \\
                                  & DisCo ($\lambda = 0.8$) & $-8.8 \pm 1.2$  & $0.6531 \pm 0.0076$ \\
                                  & MoDe ($\lambda = 0.2$)  & $-6.6 \pm 5.5$  & $0.671 \pm 0.035$ \\
        \midrule
        \bottomrule
    \end{tabular}
    \label{tab:fits}
\end{table*}

\section{Conclusion}
\label{sec:conclusion}

This paper demonstrates that the busier data-taking conditions of \lhcb in Run 3, in which bunch crossing regularly contain multiple visible \pp collisions, presents a new challenge in the form of a lifetime-correlated background.
This background must be mitigated in the topological \bquark trigger, otherwise decay-time-dependent analyses will suffer from lifetime biases, in particular a drop in efficiency at large lifetimes.
The DisCo and MoDe approaches to NN decorrelation offer ways to prevent such biases without negatively impacting their classification performance, as demonstrated for the cases of ${\BuToJpsimmK}$ and ${\BdToDPi}$ decays.
The MoDe approach fares better in these studies, with a penalty term of $\lambda = 0.2$ providing an optimal decorrelation of lifetime from the MLNNs of the topological \bquark trigger.

\begin{acknowledgements}
We would like to extend our sincere gratitude to the \lhcb Real Time Analysis project for its support in reviewing and developing this manuscript.
We are grateful to the \lhcb computing and simulation teams for producing the simulated \lhcb samples used in the development of the method and package, and in their demonstration in this manuscript.
We would also like to thank our \lhcb colleagues who have been involved in the development, implementation and validation of the methods and technologies described in this manuscript.

JA, JG, and VVG acknowledge funding from the European Union Horizon 2020 research and innovation programme, call H2020-MSCA-ITN-2020, under the Grant Agreement n. 956086.
JA and JG acknowledge the support and sponsorship of this work by the German Federal Ministry of Education and Research (BMFTR, grant no. 05H24PE2) within ErUM-FSP T04 and the Deutsche Forschungsgemeinschaft (DFG, German Research Foundation) under Germany’s Excellence Strategy – EXC 3107 – Project-ID 533766364.
AD and MG would like to express their gratitude to the Ministry of Science and Higher Education in Poland, for financial support under the contract no 2022/WK/03.
BD and MW were supported by NSF grants PHY-2019786 (The NSF AI Institute for Artificial Intelligence and Fundamental Interactions, http://iaifi.org/) and   PHY-2209181. 
CF and JED acknowledge funding from the European Union Horizon 2020 research and innovation programme, call ERC-2019-STG, under Grant Agreement n. 852642. 
MV acknowledges funding from the European Union Horizon 2020 research and innovation programme, call ERC-2019-CSG, under Grant Agreement n. 865469.
\end{acknowledgements}

\clearpage
\appendix

\section{Input features of the topological \bquark trigger models}
\label{app:features}

As described in Sec.~\ref{sec:topo}, the MLNNs of the topological \bquark trigger interpret the topological and kinematic properties of \bquark-hadron decays.
The full set of features from which the MLNNs are constructed are listed in Table~\ref{tab:features} and described below \cite{topo-chep-2023}.

The primary kinematic quantities of interest are the \pt of final state particles and reconstructed vertices, and combinations/comparisons of these, \eg, the minimum \pt of decay products associated with the 2-body vertex.
Additionally, as the reconstructed vertices typically only describe part of a decay process, the missing momentum from other products of the decay must be accounted for.
A modified invariant mass, $M_{\rm corr.}$, is used which incorporates the missing momentum transverse to the direction of flight (taken from the PV and SV), $p'_{T,{\rm missing}}$,
\begin{equation}
    M_{\rm corr.} = \sqrt{m^2 + |p'_{T{\rm, missing}}|^2} + |p_{\rm T, missing}|,
\end{equation}
for an invariant mass $m$.
Genuine \bquark-hadron decays will thus return masses closer to their true mass, allowing the NNs to suppress components such as prompt \cquark-hadron decays \cite{LHCb-PUB-2011-002}.

The topological quantities included in the NNs describe directly the spatial relation of reconstructed objects and the quality of their reconstruction:
\begin{itemize}
    \item The distance of closest approach (DOCA) of a vertex: the closest distance between the constituent particles of the vertices.
    \item The vertex fit quality, $\chi^2 / N_\mathrm{d.o.f}$.
    \item The flight-distance quality , $\chi^2_{\rm FD}$, which describes whether a vertex is significantly displaced from its PV.
    \item The impact-parameter quality, $\chi^2_{\rm IP}$, which describes how significantly far from the PV a track flies. 
\end{itemize}
\begin{table*}[htb]
    \centering
    \caption{Input features of the two- and three-body MLNNs of the topological \bquark trigger.
    The MLNNs were required to increase monotonically in features marked with \tick (\query) for all of Run 3 (for only 2022-2024) data-taking.}
    \label{tab:features}
    \begin{tabular}{ccc}
        \hline\hline
        Two-body MLNN feature & Three-body MLNN feature & Monotonic \\
        \hline
        \multicolumn{2}{c}{Min. \pt\ of 2-body vertex children} & \tick \\ 
        \multicolumn{2}{c}{Sum of \pt\ of 2-body vertex children} & \query \\ 
        \multicolumn{2}{c}{\pt\ of 2-body vertex} & \cross \\
        \multicolumn{2}{c}{$M_{\rm corr.}$ of 2-body vertex} & \cross  \\
        \multicolumn{2}{c}{$\chi^2 / N_\mathrm{d.o.f}$ of 2-body vertex} & \cross  \\
        \multicolumn{2}{c}{Max. $\chi^2_{\rm FD}$~of 2-body vertex} & \cross \\
        \multicolumn{2}{c}{Max. $\mathrm{DOCA}$ of 2-body vertex} & \cross \\
        Min. $\chi^2_{\rm IP}$~of 2-body vertex children & $\chi^2_{\rm IP}$~of 3-body vertex children & \tick \\
        Max. $\chi^2_{\rm IP}$~of 2-body vertex children & Max. $\chi^2_{\rm IP}$~of 3-body vertex children & \cross \\
        --- & $\chi^2_{\rm IP}$ of 2-body vertex & \cross \\
        --- & Min. \pt\ of 3-body vertex children & \tick \\
        --- & Sum of \pt\ of 3-body vertex children & \query \\
        --- & \pt\ of 3-body vertex & \query \\
        --- & $M_{\rm corr.}$ of 3-body vertex & \cross \\
        --- & $\chi^2 / N_\mathrm{d.o.f}$ of 3-body vertex & \cross \\
        --- & Max. $\chi^2_{\rm FD}$~of 3-body vertex & \cross \\
        --- & Max. $\mathrm{DOCA}$ of 3-body vertex & \cross \\
        \hline\hline
    \end{tabular}
\end{table*}

\section{Optimisation of decorrelation strength}
\label{app:optimisation}

The MLNNs for selecting 2- and 3-body candidates were trained with DisCo and MoDe losses, with penalty strengths according to Table~\ref{tab:lambdas}.
The strength working points were chosen according to the sensitivity of the performance on the penalty strength, \eg, including more working points for $\lambda \in \left[0.0, 0.4\right]$ for the 2-body MoDe models, since the performance varies rapidly in this range.
The training of all of the models was stable, with all converging on a minimised loss.

\begin{table*}[htb]
    \setlength{\tabcolsep}{2.8pt}
    \centering
    \caption{Values of the hyperparameter $\lambda$, determining the penalty term strength with which MLNNs were trained.}
    \label{tab:lambdas}
    \begin{tabular}{ccccccccccccccccc}
        \toprule
        Approach & Candidate & \multicolumn{15}{l}{Penalty strengths, $\lambda$} \\
        \midrule
        \multirow{2}{*}{DisCo}  & 2-body & \multirow{2}{*}{0.0} &
                                           \multirow{2}{*}{---} & 
                                           \multirow{2}{*}{---} & 
                                           \multirow{2}{*}{---} &
                                           \multirow{2}{*}{0.2} &
                                           \multirow{2}{*}{---} & 
                                           \multirow{2}{*}{---} & 
                                           \multirow{2}{*}{---} & 
                                           \multirow{2}{*}{---} &
                                           \multirow{2}{*}{0.5} & 
                                           \multirow{2}{*}{---} & 
                                           \multirow{2}{*}{---} &
                                           \multirow{2}{*}{0.8} & 
                                           \multirow{2}{*}{---} &
                                           \multirow{2}{*}{1.0} \\
                                & 3-body & \\
        \midrule
        \multirow{2}{*}{MoDe}   & 2-body & \multirow{2}{*}{0.0} & \multirow{2}{*}{0.05} & \multirow{2}{*}{0.1} & 0.15 & \multirow{2}{*}{0.2} & 0.25 & \multirow{2}{*}{0.3} & 0.35 & \multirow{2}{*}{0.4} & \multirow{2}{*}{0.5} & \multirow{2}{*}{0.6} & \multirow{2}{*}{0.7} & \multirow{2}{*}{0.8} & --- & \multirow{2}{*}{1.0} \\
                                & 3-body &  & & & --- & & --- & & --- & & & & & & 0.9 & \\
        \bottomrule
    \end{tabular}
\end{table*}

The performance of the MLNNs is assessed according to three quantities:
\begin{itemize}
    \item Precision: the fraction of true positive predictions to all positive predictions, analogous to purity of the resulting sample.
    \item Recall: the fraction of true positive predictions to the sum of true positive and false negative predictions, analogous to the selection efficiency. 
    \item Area under receiver operating characteristic (ROC) curve: as the ROC curve is the true positive rate as a function of the false positive rate, the area beneath this curve defines the classification power of the model.
    % \item F1 score: the harmonic mean of precision and recall, \ie, twice the product of precision and recall, divided by their sum, which 
\end{itemize}
An optimally performant NN would return values of 1 for each of these quantities, though in practice improvement in one is often only achievable at the expense of another.

\begin{figure*}[tb]
    \centering
    \includegraphics[width=0.45\linewidth]{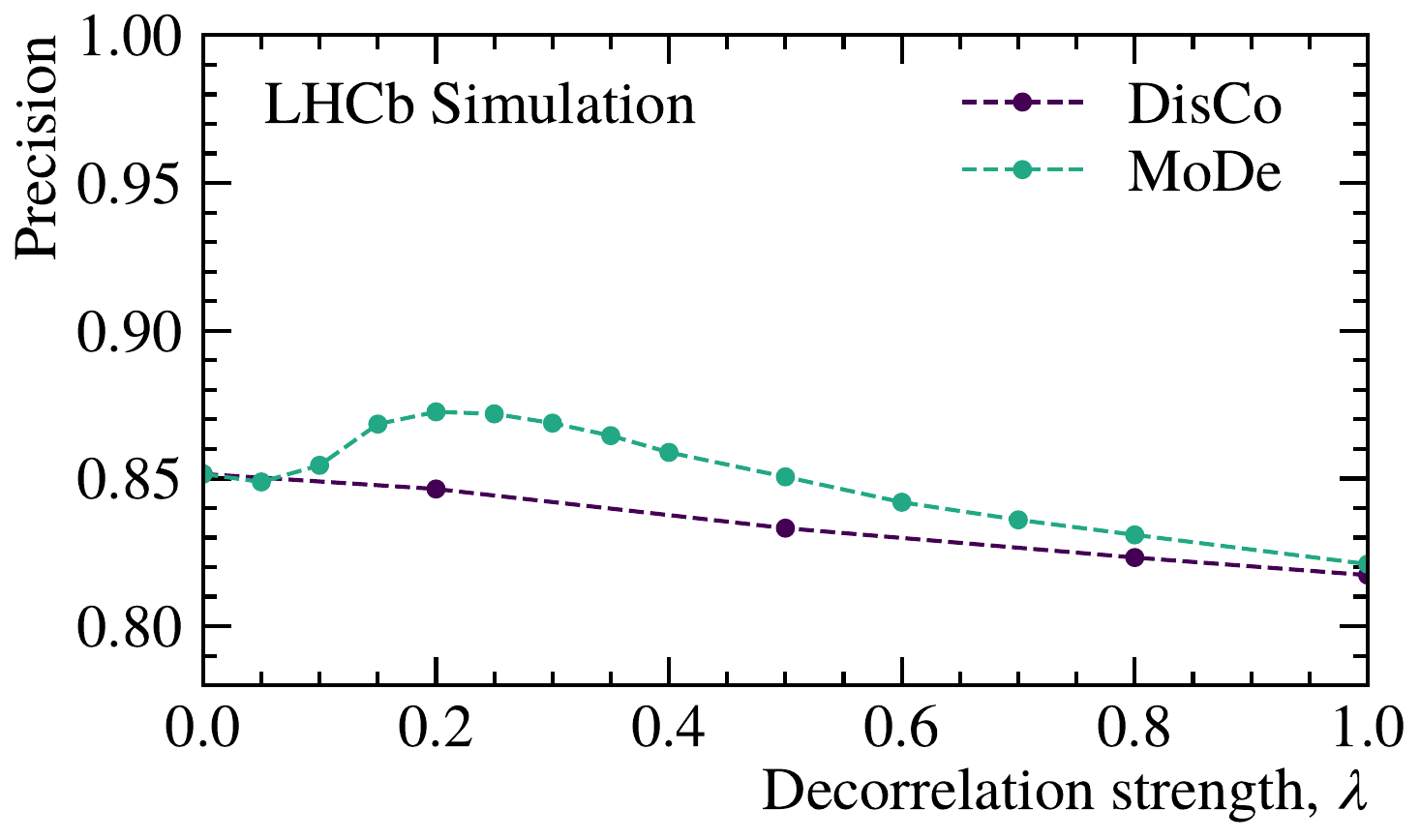}
    \includegraphics[width=0.45\linewidth]{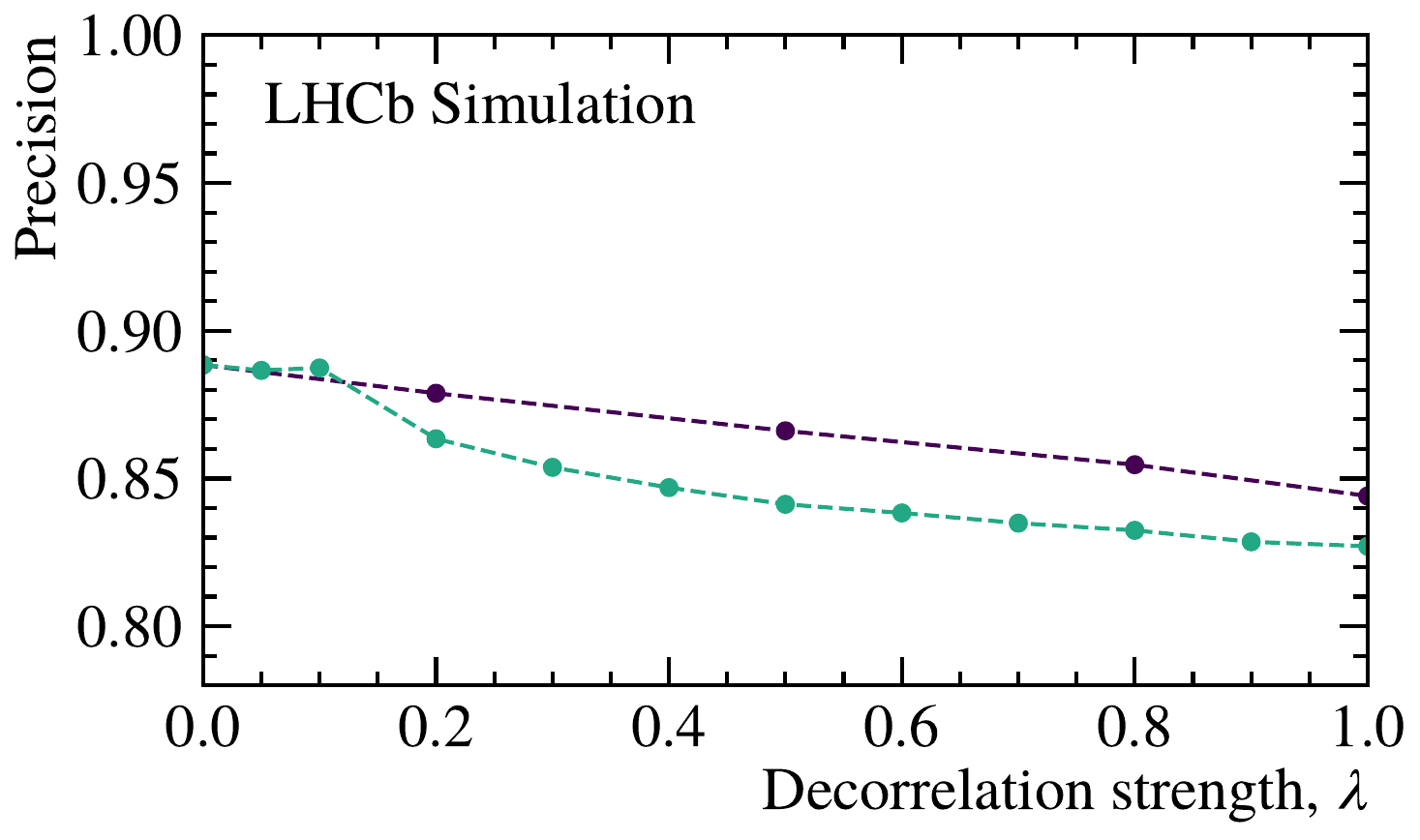}
    
    \includegraphics[width=0.45\linewidth]{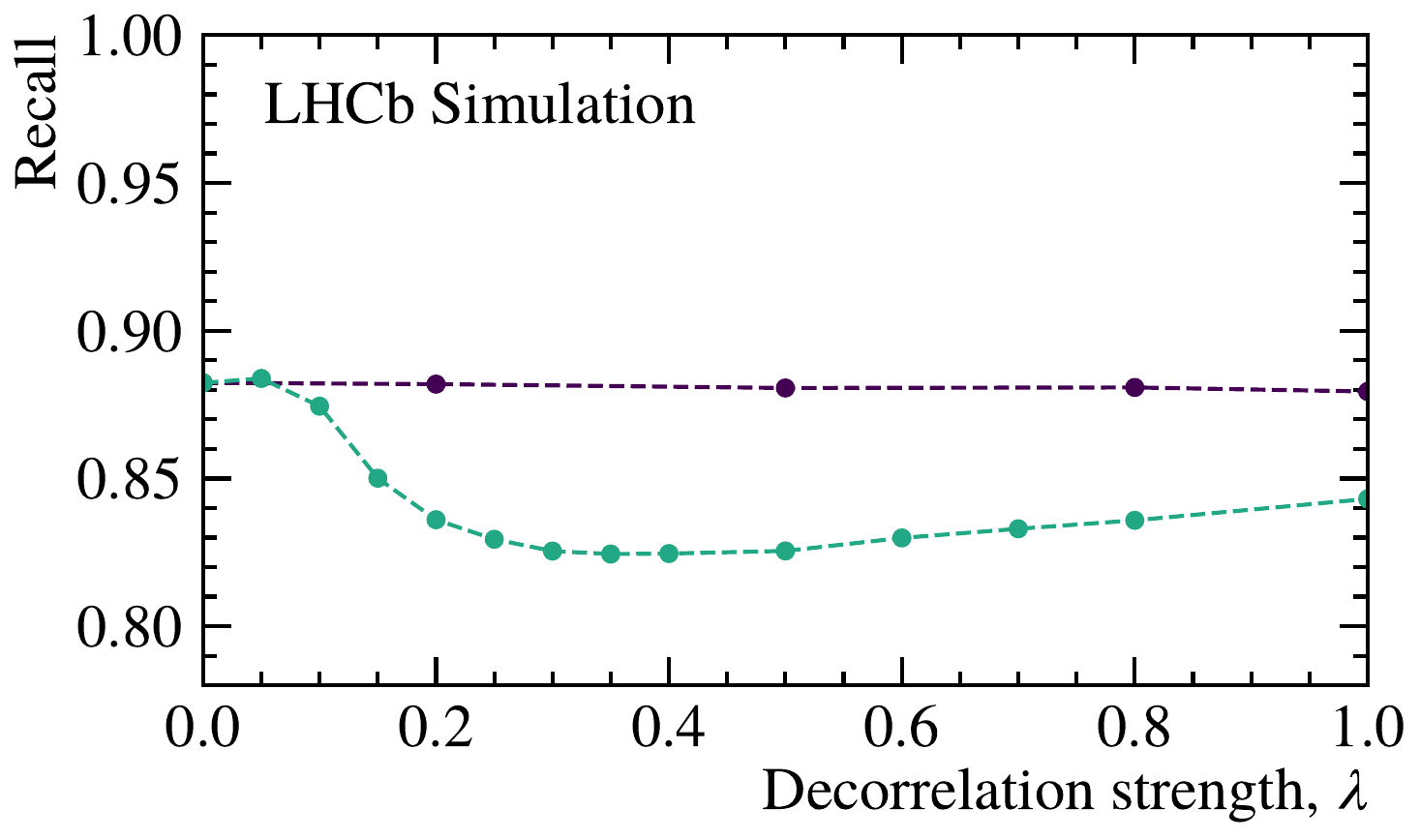}
    \includegraphics[width=0.45\linewidth]{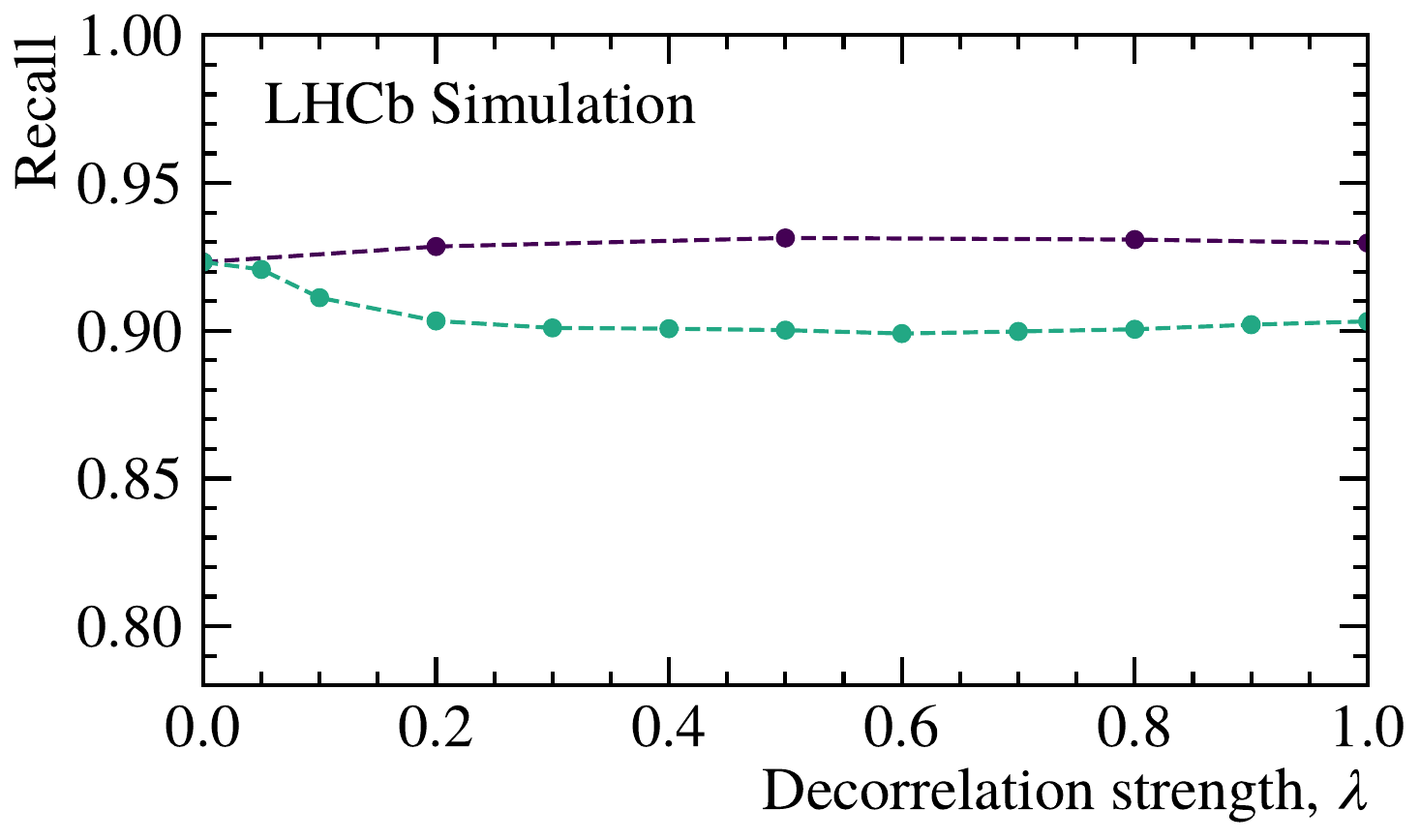}
    
    \includegraphics[width=0.45\linewidth]{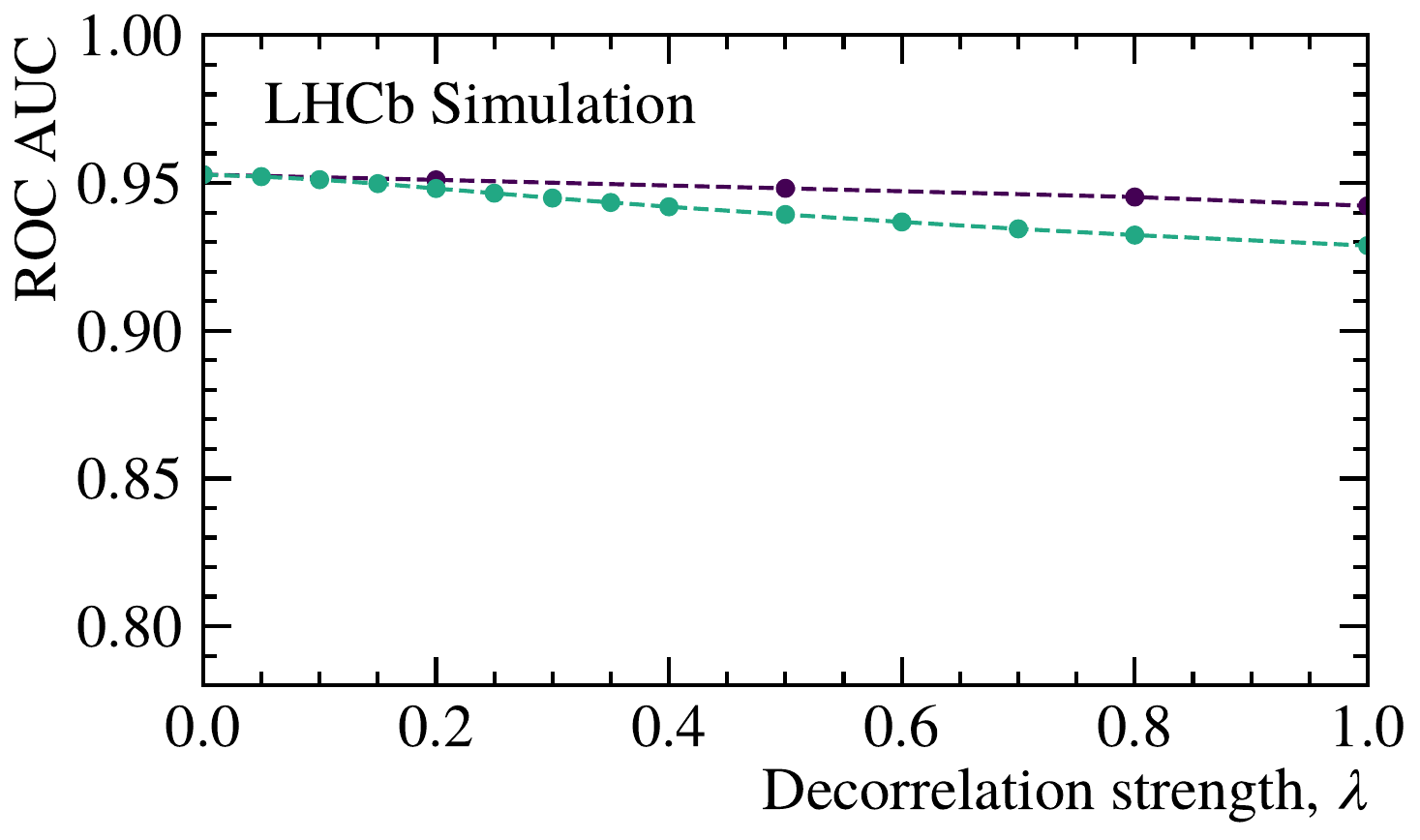}
    \includegraphics[width=0.45\linewidth]{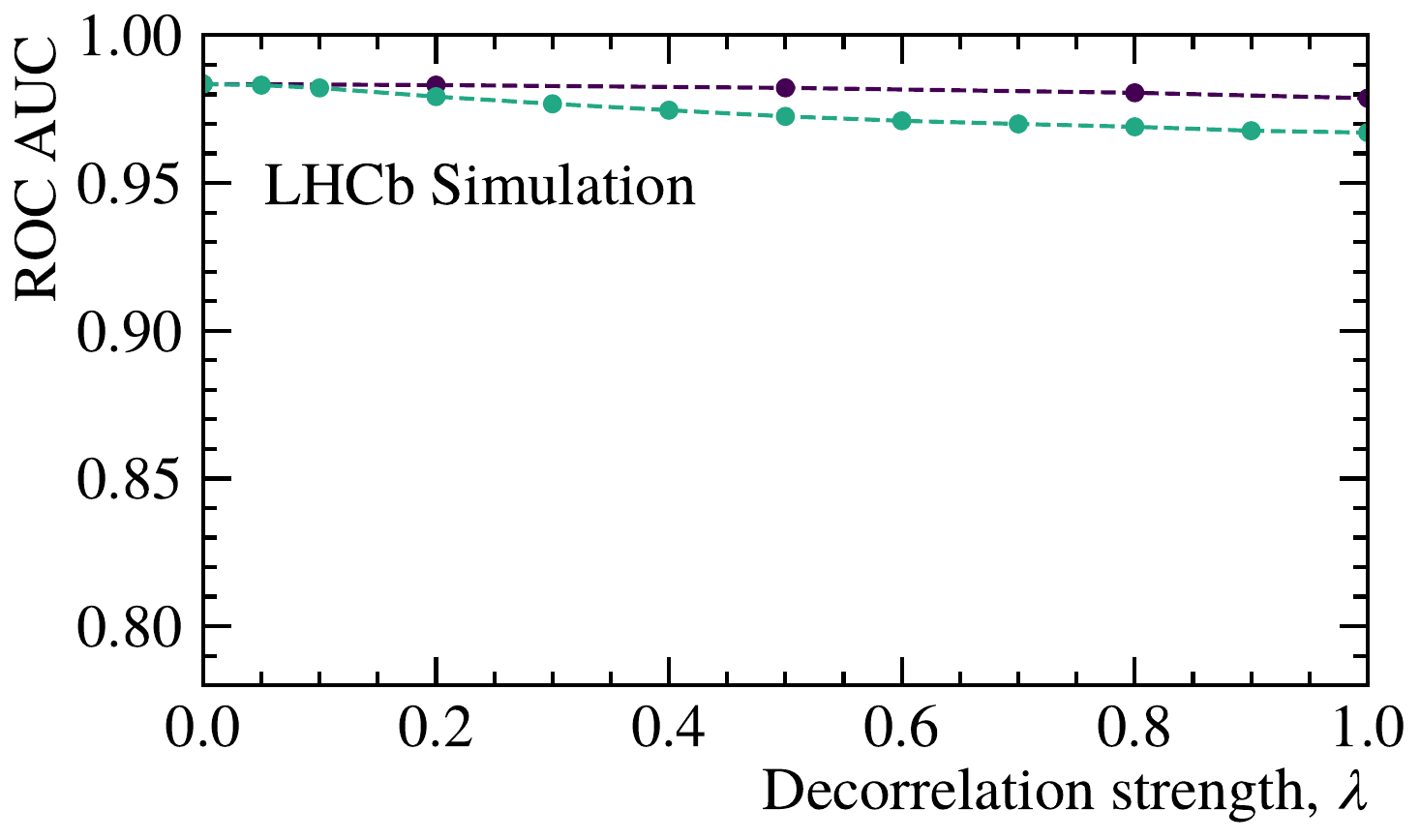}
    
    \caption{Performance of (left) 2- and (right) 3-body MLNNs, trained with DisCo and MoDe penalty terms of varying strength, $\lambda$, quantified in terms of the (top) precision, (middle) recall and (bottom) area under the ROC curve.}
    \label{fig:performance}
\end{figure*}

Each metric was evaluated for each trained model, as presented in Fig.~\ref{fig:performance}.
For the MoDe case, the 2-body model is optimally precise for $\lambda = 0.2$, for which the recall is degraded but not to a minima.
Up to this penalty strength, the area under the ROC curve is minimally degraded; however beyond $\lambda=0.2$, all three metrics decrease with increasing $\lambda$.
The 3-body model has a similar maximum precision at $\lambda = 0.1$, with significant degradation of all three metrics beyond this point.
However, as this is unlikely to significantly decorrelate the model, a strength of $\lambda=0.2$ is deemed optimal for the use case, since this provides the greatest precision, recall and area under ROC of the models beyond a strength of $\lambda = 0.1$.
For the DisCo case, an obvious optimum is not clearly visible; however, the area under the ROC curve remains minimally degraded up to $\lambda=0.8$.
Similarly, the recall improves with increasing decorrelation strength, with diminishing returns above $\lambda=0.8$.

% \end{landscape}

\bibliographystyle{spphys}       % APS-like style for physics
\bibliography{main, standard, LHCb-CONF, LHCb-DP, LHCb-PAPER, LHCb-TDR, LHCb-PUB}   % name your BibTeX data base

% % Non-BibTeX users please use
% \begin{thebibliography}{}
% %
% % and use \bibitem to create references. Consult the Instructions
% % for authors for reference list style.
% % Format for Journal Reference
% Author, Article title, Journal, Volume, page numbers (year)
% Author, Book title, page numbers. Publisher, place (year)
% \end{thebibliography}

\end{document}